\documentclass[preprint,showpacs,preprintnumbers,amsmath,amssymb]{revtex4}
\usepackage{graphicx}

\newcommand{\be}{\hbox{\normalsize e}}

\begin{document}

\title{Universal Critical Behavior in the Dicke Model}

\author{O. Casta\~nos, E. Nahmad-Achar, R. L\'opez-Pe\~na, and J. G. Hirsch}

\affiliation{Instituto de Ciencias Nucleares, Universidad Nacional
Aut\'onoma de M\'exico, Apdo. Postal 70-543 M\'exico 04510 D.F.\\ \\}


\begin{abstract}
The critical value of the atom-field coupling strength for a finite number of atoms is determined by means of both, semiclassical and exact solutions. In the semiclassical approach we use a variational procedure with coherent and symmetry-adapted states, while for the exact quantum solution the concept of fidelity is employed. These procedures allow for the determination of the phase transitions in the model, and coincide in the thermodynamic limit.
For the three cases mentioned above, universal parametric curves are obtained for the expectation values of both the first quadrature of the electromagnetic field, and the atomic relative population, as implicit functions of the atom-field coupling parameter, valid for the ground- and first-excited states.
\end{abstract}


\maketitle



\section{Introduction}

A many-body system (e.g. a cold $2$-level atomic cloud) interacting with a $1$-mode radiation field inside an optical cavity in the dipolar approximation is described by the Dicke model~\cite{dicke}. An important feature of this model is the presence of a phase transition from the normal to the superradiant behavior~\cite{hepp}. This is a collective effect involving all $N$ atoms in the sample, where the decay rate is proportional to $N^2$ instead of $N$, the expected result for independent atom emission. While this transition has been much debated in the 
literature, mainly due to the fact that it requires a very strong atom-field coupling, recent experimental results indicate that it may actually be observed in situations in which atomic transitions are replaced by transitions between momentum states~\cite{baumann,nagy}. This has brought about a renewed interest in the Dicke model, mainly for its phase transitions but also because it can present multi-partite entanglement, and can be realized in systems more widely than in the original cavity QED case (cf. e.g. \cite{garraway} and references therein).

In this work we obtain the phase transition of the Dicke model for a {\it finite} number $N$ of atoms, via $3$ different methods: i) through a numerical diagonalization of the Hamiltonian and the use of the fidelity between neighbouring states; ii) through variational test states that are a direct product of coherent Heisenberg-Weyl ($HW(1)$)-states (for the electromagnetic field), and $SU(2)$-states (for the atomic field)~\cite{papercorto,paperextenso}; and iii) through the use of projection operators on the coherent states in (ii) to obtain states which obey the parity symmetry in the total excitation number, present in the Hamiltonian. All these procedures coincide in the thermodynamic limit $N\rightarrow\infty$. We show that, for a finite number of atoms, the {\it symmetry-adapted} states (iii) constitute a much better approximation to the exact quantum ground- and first excited-states. (Needless to say, the symmetry is also exploited in the numerical diagonalization of the Hamiltonian.)

For the three cases mentioned above, universal parametric curves are obtained for the expectation values of both the first quadrature of the electromagnetic field, and the atomic relative population, as functions of the atom-field coupling parameter, valid for the ground- and first-excited states.

\section{Phase transitions and universal behavior}

The Dicke Hamiltonian involves the collective interaction of $N$
two-level atoms, with energy separation $\hbar\tilde{\omega}_{A}$, with
a one-mode radiation field of frequency $\tilde{\omega}_{F}$, in the
long wavelength limit. It has the form
   \begin{equation}
      H_{D}=\hat{a}^{\dagger}\hat{a}+\omega_{A}\hat{J}_{z}
      +\frac{\gamma}{\sqrt{N}}\left(\hat{a}^{\dagger}+\hat{a}\right)
      \left(\hat{J}_{+}+\hat{J}_{-}\right)\ ,
      \label{eq001}
   \end{equation}
where $\omega_{A} = \tilde{\omega}_{A} / \tilde{\omega}_{F}$ and $\gamma$, the coupling parameter between the matter and field, are given in units of the frequency of the field, and we have taken $\hbar = 1$. The operators
$\hat{a}^{\dagger},\,\hat{a}$ denote the one-mode creation and annihilation
photon operators; $\hat{J}_{z}$ the atomic relative population operator;
and $\hat{J}_{\pm}$ the atomic transition operators.

The energy surface is found by taking the 
expectation value of the Dicke Hamiltonian  with respect to
the tensorial product of coherent states for the $HW(1)$ and $SU(2)$ groups
$\vert\alpha\rangle\otimes\vert\zeta\rangle\ $, given
by~\cite{hecht,gilmore1972}
   \begin{eqnarray*}
      \vert\alpha\rangle&=&\exp\left(-\left|\alpha\right|^{2}/2\right)\,
      \sum_{\nu=0}^{\infty} \frac{\alpha^{\nu}}{\sqrt{\nu!}}
      \,\vert\nu\rangle\ ,\\
      \vert\zeta\rangle&=&\frac{1}{\left(1+
      \left|\zeta\right|^{2}\right)^{j}}\,\sum_{m=-j}^{j}
      \binom{2j}{j+m}^{1/2}\,\zeta^{j+m}\,\vert j,\,m\rangle\ ,
   \end{eqnarray*}
where the parameters $\alpha$ and $\zeta$ are complex numbers.

Straightforward calculation leads to~\cite{ocasta1,paperextenso}
   \begin{eqnarray}
      {\cal H}(\alpha,\,\zeta)&=&\frac{1}{2}\left(p^{2}+q^{2}\right)-j \,
      \omega_{A}\,\cos\theta+2\sqrt{j}\gamma\,q\,\sin\theta\,
      \cos\phi\ .
      \label{eq003}
   \end{eqnarray}
In this expression we use the harmonic oscillator realization
for the field variables and the stereographic projection for the angular
momentum parameters,
	\begin{equation}
		\alpha=\frac{1}{\sqrt{2}}\left(q+i\,p\right)\ ,
                \qquad\zeta=\be^{-i\,\phi}\,\tan\frac{\theta}{2}\ ,
		\label{dzeta}
	\end{equation}
where $(q,p)$ correspond to the expectation values of the
quadratures of the field, and $(\theta,\phi)$ determine a
point on the Bloch sphere.

\begin{figure}[h]
\scalebox{0.7}{\includegraphics{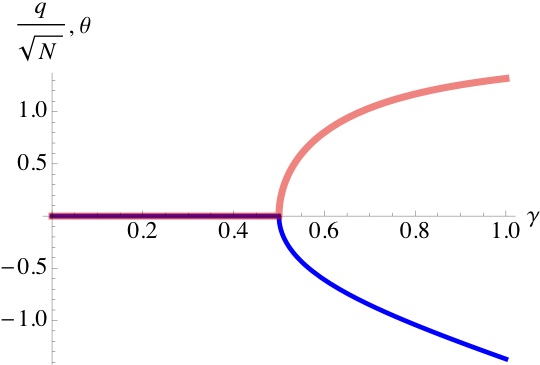}}
\caption{\label{fig1}
(Color Online.) Critical values $\frac{q_c}{\sqrt{N}}$ (lower, blue) and $\theta_c$ (upper, red) as functions of the interaction strength $\gamma$, for $\omega_{A}=1$.}
\end{figure}

Using the Ritz variational principle one finds the best variational approximation to the ground state energy of the system and its corresponding eigenstate.
The minima and degenerate critical points are obtained by means of the
catastrophe formalism. For the coherent states we find that when $\gamma_{c}^{2}=\omega_{A}/4$ the
critical points degenerate; for this critical value of the field-matter
coupling, a phase transition takes place from the normal to the superradiant
behavior of the atoms~\cite{gilmore3}. The critical points which minimize ${\cal H}$ are given by
	\begin{equation}\label{criticos}
		\begin{array}{lllll}
		\theta_{c}=0\,,& q_{c}=0\, ,&p_{c}=0\, ,&\hbox{
                for } \vert\gamma\vert < \gamma_c \, ,\\
		\theta_{c}=\arccos(\gamma_c/\gamma)^{2}
		\, ,&q_{c}=-2\,\sqrt{j}\,\gamma\,
		\sqrt{1-(\gamma_c/\gamma)^{4}} 
		\cos{\phi_c}\, ,&p_{c}=0\, ,&\hbox{ for
                }\vert\gamma\vert > \gamma_c\ ,
		\end{array} 
	\end{equation}
where $\phi_{c}=0,\,\pi$.
Fig.~\ref{fig1} shows $q_c/\sqrt{N}$ and $\theta_c$ as a function of $\gamma$, for $\phi_c=0$, $\omega_{A}=1$ and $N=20$ atoms. The phase transition is clearly observed at $\gamma=\gamma_c=1/2$. If we were to use the critical point $\phi_c=\pi$, the curve describing the quadrature of the radiation field would change sign.

\begin{figure}[h]
\scalebox{0.7}{\includegraphics{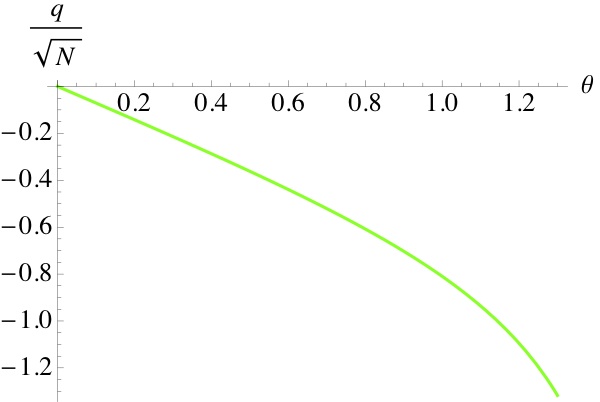}}
\caption{\label{fig2}
Universal curve $q_c/\sqrt{N}$ versus $\theta_c$, for $\omega_{A}=1$.}
\end{figure}

From the relations in Eq.(\ref{criticos}) we obtain, in the superradiant region,
	\begin{equation}\label{qtheta}
		\frac{q_c}{\sqrt{N}}= - \sqrt{\omega_A} \frac{\sin\theta_c}{\sqrt{2\,\cos\theta_c}}\cos\phi_c\ .
	\end{equation}
This is plotted in Fig.~\ref{fig2}. Notice that the normal regime is described only by the origin $(q_c=0,\, \theta_c=0)$, and that the curve is valid for {\it all} $\vert\gamma\vert > \gamma_c$ (as this parameter drops out from Eq.(\ref{qtheta})). It is also valid for {\it any} number of atoms. As we will show below, the same curve is obtained for symmetry-adapted variational states as well as for exact quantum solutions; in this sense the curve is universal. Furthermore, the first-excited states fall along the same path.

The Hamiltonian (\ref{eq001}) shows a parity symmetry given by
	\begin{equation}
		[e^{i \pi \Lambda},\, H] = 0
	\end{equation}
with $\Lambda = a^\dagger a + J_z + \sqrt{J^2 + \frac{1}{2}} - \frac{1}{2}$, the {\it total excitation number operator}. This allows for the classification of its eigenstates in terms of the parity of the eigenvalues 
$\lambda = \nu + m + j$ of $\Lambda$, where $\nu$, $m$, and $j(j+1)$ are eigenvalues of $a^\dagger a$, $J_z$, and $J^2$, respectively. This allows us to build, through the projection operators $P_{\pm}=\frac{1}{2}\left(I \pm e^{i\,\pi\,\Lambda}\right)$, {\it symmetry-adapted} coherent states (SAS) that preserve the symmetry of the Dicke
Hamiltonian, i.e., have a definite parity of the total number of excitations. These are given by~\cite{paperextenso}
	\begin{equation}
		|\alpha,\,\zeta \rangle_{\pm}={\cal N}_{\pm}\Big(
		\vert\alpha\rangle\otimes\vert\zeta\rangle\pm\
		\vert-\alpha\rangle\otimes\vert-\zeta\rangle\Big)\ ,
		\label{eq013}
	\end{equation}
where the normalization factors ${\cal N}_{\pm}$ are
       \begin{equation}
          {\cal N}_{\pm}^{-2}=2\,\left(1\pm\exp\left(-2\,|\alpha|^{2}
          \right)\left(\frac{1-|\zeta|^2}{1+|\zeta|^2}\right)^{N}\right)\ .
          \label{eq014}
       \end{equation}

The energy surface associated to SAS is given by~\cite{paperextenso}
	\begin{eqnarray}
	\langle H \rangle_\pm &=& \pm \frac{1}{2} \left(p^2+q^2\right)
	\left\{1-\frac{2}{1 \pm e^{\pm(p^2+q^2)} (\cos\theta)^{\mp
            N}}\right\}\nonumber \\
	&-&\frac{N}{2} \, \omega_{A}
	 \left\{(\cos \theta)^{\pm 1} 
	 \pm \frac{\tan^2\theta \, \cos\theta }{1 \pm e^{\pm(p^2+q^2)} 
	(\cos\theta)^{\mp N} }\right\} \nonumber \\
	&+& \sqrt{2 \, N} \, \gamma  \left\{\frac{\pm p \, \tan\theta \, 
	\sin\phi + q \, e^{p^2+q^2}
	\sin\theta \, \cos \phi \, (\cos\theta)^{-N} }{
	e^{p^2+q^2} (\cos\theta)^{-N}  \pm 1 } \right\} \, .
	\label{symad}
	\end{eqnarray}
It is straightforward to show that in the thermodynamic limit both surfaces $\langle H \rangle_+$ and $\langle H \rangle_-$ reduce to Eq.(\ref{eq003}).

\begin{figure}[h]
\scalebox{0.7}{\includegraphics{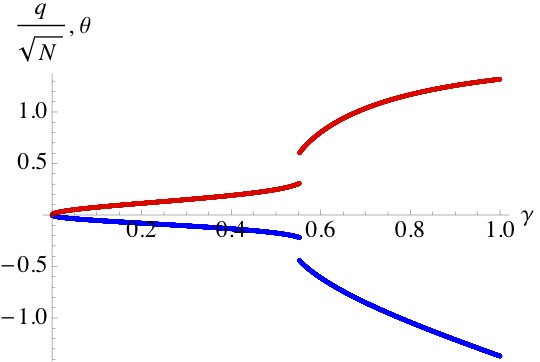}}\qquad
\scalebox{0.7}{\includegraphics{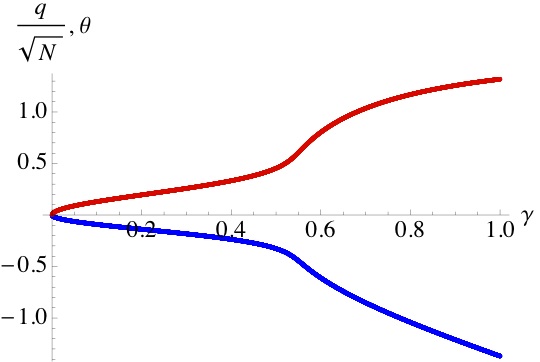}}
\caption{\label{fig3}
(Color online.) Critical values of $q/\sqrt{N}$ (lower, blue), and $\theta$ (upper, red) as functions of the interaction strength $\gamma$, for $\phi_c=0$, $\omega_{A}=1$ and $N=20$ atoms.  Left: even SAS (ground state); right: odd SAS (first excited state).}
\end{figure}

For a given number of particles $N$ and a fixed coupling parameter $\gamma$,  we determine the values of $q_c$, $p_c$, $\theta_c$,
$\phi_c$, where the energy surface for the SAS states has its minimum value, in the same manner as done for the coherent states (CS). The minima of the SAS energy surface occurs for $p_c=0$ and $\phi_c=0,\,\pi$. The values of $q_c$ and $\theta_c$, which now depend strongly on $N$, are found numerically and they are plotted as a function of $\gamma$ in Fig.~\ref{fig3} for the even and odd variational states.

\begin{figure}[h]
\scalebox{0.5}{\includegraphics{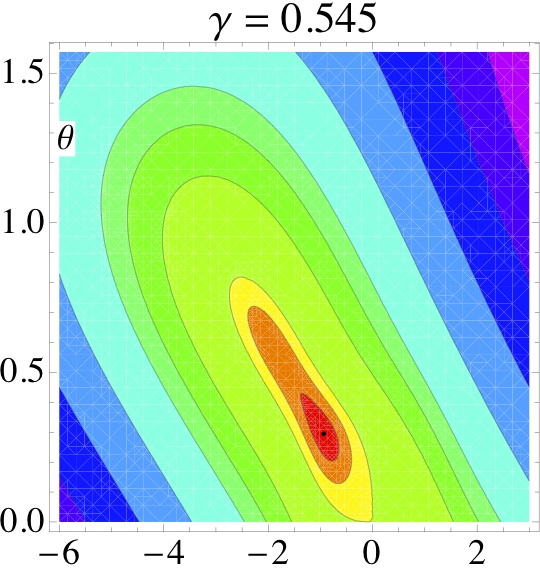}}\quad
\scalebox{0.5}{\includegraphics{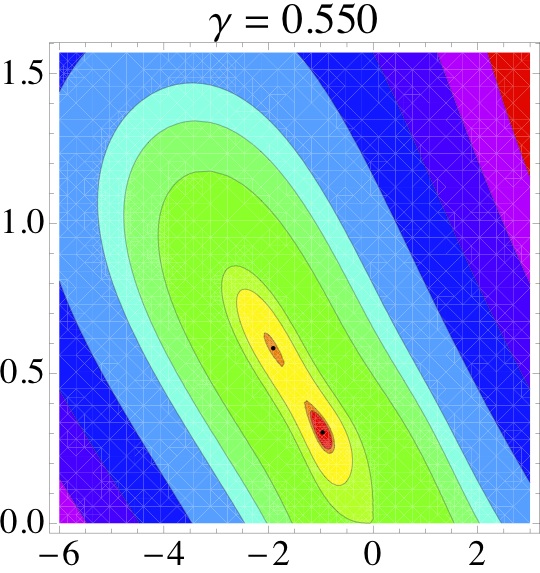}}\quad
\scalebox{0.5}{\includegraphics{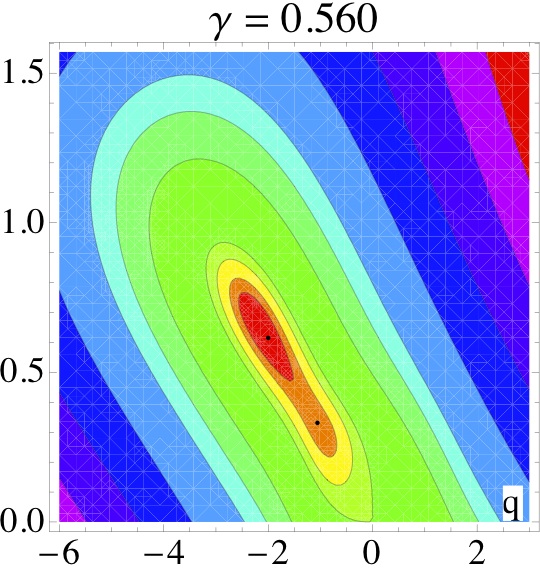}}
\caption{\label{fig4}
Contours for the even energy surface $\langle H \rangle_{+}$ showing the jump of the global minimum as the interaction strength $\gamma$
crosses its critical value at the phase transition. Plotted are the critical values of $q$ and $\theta$, for $N=20$ particles and $\omega_A=1$. $\{E_{min},\, q_{min},\, \theta_{min}\}$ are, from left to right, \{-10.1887,\,-0.935972,\,0.29446\}, \{-10.1963,\,-0.964931,\,0.30329\} and \{-10.2559,\,-2.0,\,0.615064\}.}
\end{figure}

The discontinuity in the phase space variable for the even states marks the phase transition. This phase transition arises from a jump of the global minimum in the even energy surface $\langle H \rangle_{+}$ between two local minima, as $\gamma$ varies; this can be seen in Fig.~\ref{fig4} where, for three values of $\gamma$ crossing the discontinuity, the contour energy levels are plotted as functions of $q$ and $\theta$, for $\phi_c=0$.  The jump occurs at $\gamma=0.553$ for $N=20$. For the odd state, the appearance of a phase transition manifests itself through a change in the concavity of the associated curves for $q_c$ and $\theta_c$.

\begin{figure}[h]
\scalebox{0.7}{\includegraphics{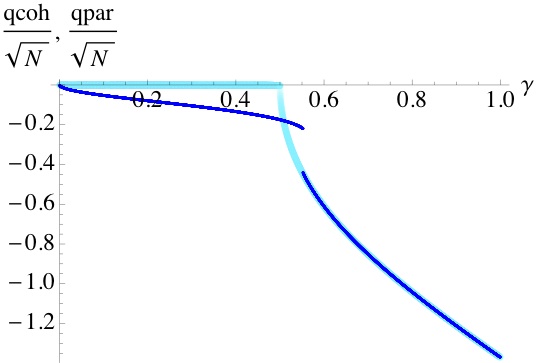}}\qquad\qquad
\scalebox{0.7}{\includegraphics{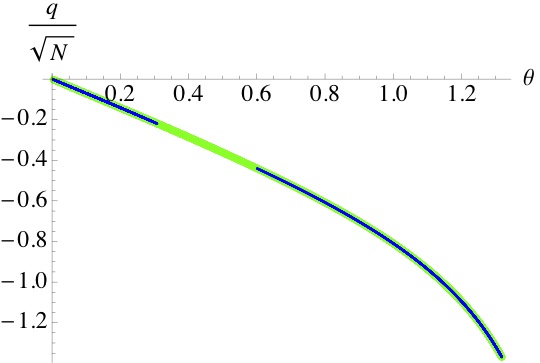}}
\caption{\label{fig5}
{Color online.} Left: $q_c/\sqrt{N}$ as a function of $\gamma$ for the coherent (continuous, cyan) and even SAS (discontinuous, blue) estimations of the ground state. Beyond the phase transition both curves coincide. Right: $q_c/\sqrt{N}$ as a function of $\theta_c$ for the CS and odd SAS states (continuous, green) and even SAS state (discontinuous,blue).}
\end{figure}

It is useful to compare the minimum points $q_c$ associated to the energy surfaces of the CS and the even SAS (cf. Fig.~\ref{fig5}). The quadrature $q_c$ shows a discontinuity for the latter at the phase transition, as a function of the coupling parameter $\gamma$. This translates into a forbidden range of values for $\theta_c$ ($0.3\le\theta\le 0.6$) when $q_c$ is plotted against it. Note that, in the allowed $\theta$-region, the functions $q_c(\theta_c)$ associated to the energy surfaces of the CS and even SAS fall along the same curve. The odd SAS state is also plotted, falls along the same curve, has no forbidden regions, and is indistinguishable from that of the coherent state.

%

Since the $q_c$ vs. $\theta_c$ seems to show a universal behavior, it is convenient to study the exact quantum solution. We propose the following correspondence between the critical $(q_c,\theta_c)$ parameters of the classical phase space and the quantum operators:
\begin{equation}
q_c  \longrightarrow  \mp \sqrt{2 \langle a^{\dagger}a\rangle} \, , \qquad \theta_c \longrightarrow \arccos\left({ \frac{-\langle J_{z}\rangle}{j}}\right) \, ,
\label{corresp}
\end{equation}
where $j=N/2$ and the $\mp$ corresponds to $\phi_c=0, \, \pi$. As the system is not integrable, we solve numerically the Hamiltonian eigenvalue equation for the ground- and first-excited states, separating the even- and odd-parity cases.

\begin{figure}[h]
\scalebox{0.25}{\includegraphics{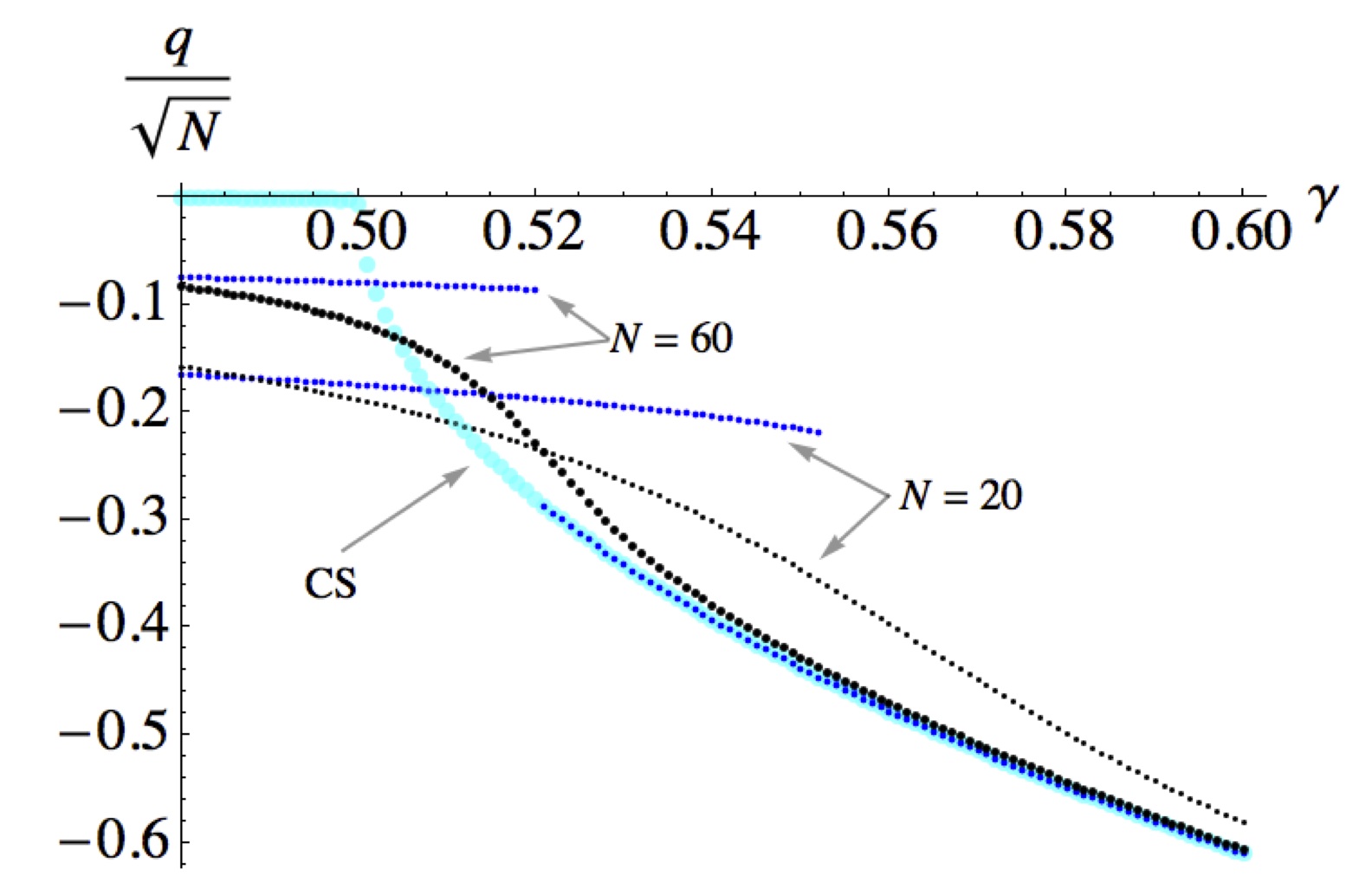}}\qquad
\scalebox{0.84}{\includegraphics{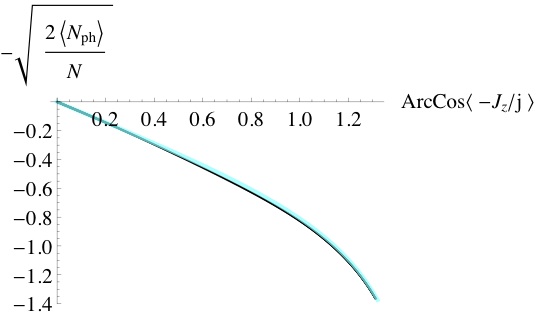}}
\caption{\label{fig6}
Left: Dependence with $\gamma$ of the quadrature (or expectation value of the number operator), for $N=20$ (lower continuous, gray dotted line) and $N=60$ (upper continuous, black dotted line), compared with those for the SAS state at the same values of $N$ (discontinuous, dotted blue lines) and with the result for the coherent state (upper continuous cyan line). Right: Quantum state $q_c$ vs. $\theta_c$ ($\sqrt{\langle a^{\dagger}a\rangle/j}$ vs. $\arccos(\langle J_{z}\rangle/j)$) behavior, falling exactly on the universal curve together with the CS and SAS states.}
\end{figure}

Fig.~\ref{fig6} shows the results. The figure on the left shows, in the vicinity of $\gamma_c$, the dependence with $\gamma$ of the quadrature (or expectation value of the number operator), for $N=20$ (lower continuous, gray dotted line) and $N=60$ (upper continuous, black dotted line). These are compared with those for the SAS state at the same values of $N$ (discontinuous, dotted blue lines) and with the result for the coherent state (upper continuous cyan line). Note that the quantum and SAS solutions agree perfectly at $\langle N_{ph}\rangle = q_c = 0$, and approximate very well beyond the phase transition. While the SAS plot is discontinuous at the phase transition, the quantum solution shows an inflection point.  The CS agrees beyond $\gamma = \gamma_c$, but fails to reproduce the results well enough in the normal region. As $N$ increases, $\gamma_c \rightarrow 0.5$ (the value for the coherent state), and the jump in the SAS curves becomes smaller, the curves approaching more and more its quantum counterparts. The figure on the right shows the quantum state $q_c$ vs. $\theta_c$ ($\sqrt{\langle a^{\dagger}a\rangle/j}$ vs. $\arccos(\langle -J_{z}\rangle/j)$) behavior for $N=20$, falling indistiguishably on the universal curve together with the CS and SAS states.

\begin{figure}[h]
\scalebox{0.8}{\includegraphics{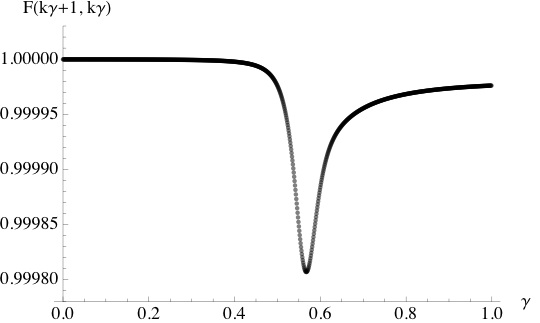}}
\caption{\label{fig7}
Quantum phase transition determined through the fidelity of $\gamma$-neighbouring states. For $N=20$ and $\omega_A=1$ the quantum
phase transition occurs at $\gamma=0.567\,\pm\,0.001$.}
\end{figure}

The quantum phase transition was found through the fidelity between the state at $\gamma + \delta\gamma$ and the state at $\gamma$~\cite{zanardi}
\begin{equation}
\mathcal{F} = |\langle\psi(\gamma)\,|\,\psi(\gamma+\delta\gamma)\rangle|^2\,,
\end{equation}
which at the transition acquires its minimum value, by taking $\delta\gamma=0.001$; for $N=20$ we find the transition at $\gamma_c=0.567\,\pm\,0.001$, comparing very well with that obtained for the SAS states, even for this small $N$ (cf. Fig.~\ref{fig7}). Another method is to use its second derivative, the so-called {\it fidelity susceptibility}, which, besides pinpointing the phase transition, it allows the determination of the scaling behavior with respect to the number of particles~\cite{zanardi, gu, ocasta3}.

\section{Conclusions}

A universal curve for all values of the coupling parameter $\gamma$ and {\it any} number of atoms $N$ is obtained for the phase space variables $q$ versus $\theta$, valid for the ground- and first-excited states. The universal character is irrespective of whether one uses coherent CS states, symmetry-adapted SAS states, or the exact quantum solution, even though these quantities are arrived at via very different methods. The gap in this curve present for the SAS states closes in the thermodynamic limit.

The critical value $\gamma_c$ at which the phase transition occurs is calculated, for the exact solution, by using the fidelity between neighboring states in the coupling strength. This agrees very well, though not identically, with the critical value given by the SAS states. For instance, in the case $N=20$ we find $\gamma_c \vert_{SAS}=0.553$ while $\gamma_c \vert_{quant}=0.567$. These values become even closer as $N$ increases. This will be studied thoroughly by means of the fidelity susceptibility elsewhere, together with the properties of the ground- and first-excited states {\it at} the phase transition, for a {\it finite} number of atoms. These results are particularly relevant in atomic physics and quantum optics using superconducting $q$-bits~\cite{nori}, where having semi-analytical solutions for small values of $N$ is very useful.

While for the coherent states, or mean-field procedure, the phase transition is determined properly only in the thermodynamic limit, for our SAS variational states the behavior of both, the ground- and first-excited states, are determined by means of jumps in the phase space variables as functions of the coupling parameter between the electromagnetic field and the matter, and may be calculated for a finite number of particles. As expected, in the thermodynamic limit all solutions converge to the mean field result.

\section*{Acknowledgements}

This work was supported by CONACyT (under project 101541) and DGAPA-UNAM (under project IN102811).




\begin{thebibliography}{9}
	
\bibitem{dicke}
R. H. Dicke, Phys. Rev. {\bf 93}, 99 (1954).

\bibitem{hepp}
K. Hepp and E. H. Lieb, Phys. Rev. {\bf A 8}, 2517 (1973).

\bibitem{baumann}
K. Baumann, C. Guerlin, F. Brennecke, and T. Esslinger, Nature {\bf 464}, 1301 (2010).

\bibitem{nagy}
D. Nagy, G. Konya, G. Szirmai, and P. Domokos, Phys. Rev. Lett. {\bf104}, 130401 (2010)

\bibitem{garraway}
B.M. Garraway, Phil. Trans. R. Soc. A {\bf 369}, 1137 (2011).

\bibitem{papercorto}
O. Casta\~nos, E. Nahmad-Achar, R. L\'opez-Pe\~na, and J.G. Hirsch, Phys. Rev. A {\it Rapid Comm.} {\bf 83}, 051601(R) (2011).

\bibitem{paperextenso}
O. Casta\~nos, E. Nahmad-Achar, R. L\'opez-Pe\~na, and J.G. Hirsch, Phys. Rev. A {\bf 84}, 013819 (2011).

\bibitem{hecht}
K.T. Hecht, {\sl The Vector Coherent State Method and Its Applications
to Problems of Higher Symmetries} (Springer, Berlin 1987).

\bibitem{gilmore1972}
F.T. Arecchi, E. Courtens, R. Gilmore, and H. Thomas, Phys. Rev. 
{\bf A6}, 2211 (1972).

\bibitem{ocasta1}
O. Casta\~nos, E. Nahmad-Achar, R. L\'opez-Pe\~na, and J.G. Hirsch in {\it Symmetries in Nature} 
AIP Conference Proceedings {\bf 1323}, p. 40-59 
Eds. L. Benet, P.O. Hess, J.M. Torres, K.B. Wolf 
(Melville, New York, 2010).

\bibitem{gilmore3}
R. Gilmore, {\sl Catastrophe Theory for scientists and engineers},
(Wiley, New York, 1981).

\bibitem{zanardi}
P. Zanardi and N. Paunkovic, Phys. Rev. E {\bf 74}, 031123 (2006).

\bibitem{gu}
S-J. Gu, Int. J. Mod. Phys. B {\bf 24}, 4371 (2010).

\bibitem{ocasta3}
O. Casta\~nos, R. L\'opez-Pe\~na,  E. Nahmad-Achar, and J.G. Hirsch in {\it XXXV Symposium on Nuclear Physics}
IOP Journal of Physics: Conference Series (2012, to be published).

\bibitem{nori}
J.Q. You and F. Nori, Nature {\bf 474}, 589 (2011); J. Clarke and F.K. Wilhelm, Nature {\bf 453}, 1031 (2008).




%
%
%
%
%

%
%
%
%
%

%
%
%
%
%
%
%
%
%
%
%
%
%
%
%


\end{thebibliography}
\end{document}